\documentclass[aip,preprint]{revtex4-1}

\usepackage{graphicx}
\usepackage{dcolumn}
\usepackage{bm}
\usepackage{soul}
\usepackage{color}
\usepackage{amsmath}
\usepackage[colorlinks,linkcolor=black,citecolor=black,urlcolor=black]{hyperref}

\newcommand{\Lancaster}{Department of Physics, Lancaster University, Lancaster, LA1 4YB, United Kingdom.}
\newcommand{\Aivon}{Aivon Oy, Espoo, Finland.}
\newcommand{\VTT}{VTT Technical Research Centre of Finland, P.O. Box 1000, 02044 VTT Espoo, Finland.}

\newcommand{\Te}{T_\mathrm{e}}
\newcommand{\Tp}{T_\mathrm{p}}
\newcommand{\Tmxc}{T_\mathrm{mxc}}
\newcommand{\Vdc}{V_\mathrm{DC}}
\newcommand{\Idc}{I_\mathrm{DC}}
\newcommand{\Iac}{I_\mathrm{AC}}
\newcommand{\mK}{\,\mathrm{mK}}

\begin{document}

\title{Nanoelectronic thermometers optimised for sub-10 millikelvin operation}

\author{J.~R.~Prance}
\email[Electronic mail: ]{j.prance@lancaster.ac.uk}
\affiliation{\Lancaster}
\author{D.~I.~Bradley}\affiliation{\Lancaster}
\author{R.~E.~George}\affiliation{\Lancaster}
\author{R.~P.~Haley}\affiliation{\Lancaster}
\author{Yu.~A.~Pashkin}\affiliation{\Lancaster}
\author{M.~Sarsby}\affiliation{\Lancaster}

\author{J.~Penttil{\"a}}\affiliation{\Aivon}
\author{L.~Roschier}
\email[Electronic mail: ]{leif.roschier@iki.fi}
\affiliation{\Aivon}

\author{D.~Gunnarsson}\affiliation{\VTT}
\author{H.~Heikkinen}\affiliation{\VTT}
\author{M.~Prunnila}
\email[Electronic mail: ]{mika.prunnila@vtt.fi}
\affiliation{\VTT}

\begin{abstract}
We report the cooling of electrons in nanoelectronic Coulomb blockade thermometers below $4\mK$. Above $7\mK$ the devices are in good thermal contact with the environment, well isolated from electrical noise, and not susceptible to self-heating. This is attributed to an optimised design that incorporates cooling fins with a high electron-phonon coupling and on-chip electronic filters, combined with a low-noise electronic measurement setup. Below $7\mK$ the electron temperature is seen to diverge from the ambient temperature. By immersing a Coulomb Blockade Thermometer in the $^3$He/$^4$He refrigerant of a dilution refrigerator, we measure a lowest electron temperature of $3.7\mK$.
\end{abstract}

\maketitle

Cooling nanoelectronic structures to millikelvin temperatures presents extreme challenges in maintaining thermal contact between the electrons in the device and an external cold bath \cite{Giazotto2006}. It is typically found that the electrons in a nanoelectronic device cooled even to $\sim 10\mK$ are significantly overheated. Understanding how to obtain and measure electron temperatures approaching $1\mK$ will open a new regime for studying nanoelectronics, and pave the way towards pioneering sub-millikelvin techniques \footnote{\href{http://www.microkelvin.eu/}{European FP7 Programme MICROKELVIN Project (http://www.microkelvin.eu/)}}. This would benefit, for example, investigations of the fractional quantum Hall effect in 2D electron gases\cite{Pan1999,Samkharadze2011}, and solid-state quantum technologies including superconducting and semiconducting qubits. To access these temperatures, one must minimise parasitic heating and internal Joule heating, maximise the coupling to cold contact wires and phonons in the host lattice, all the while overcoming the decrease in electron-phonon coupling and electrical heat conduction as temperatures drop.

Here we study Coulomb Blockade Thermometers (CBTs) that have been designed to operate significantly below $10\mK$ and demonstrate cooling of electrons in a nanoelectronic device below $4\mK$. This was achieved using CBTs as a diagnostic tool to quantify and optimise the thermal environment of the electrons in the device. A CBT consists of  an array of Coulomb-blockaded metallic islands connected by tunnel junctions whose conductance is temperature dependent\cite{Pekola1994,Meschke2011}. CBTs typically function over a decade of temperature and have previously been demonstrated to work at temperatures as low as $7.5\mK$\cite{Casparis2012}. Perhaps most importantly, they can be viewed as a primary thermometer of their internal electron temperature. 

\begin{figure}
\includegraphics{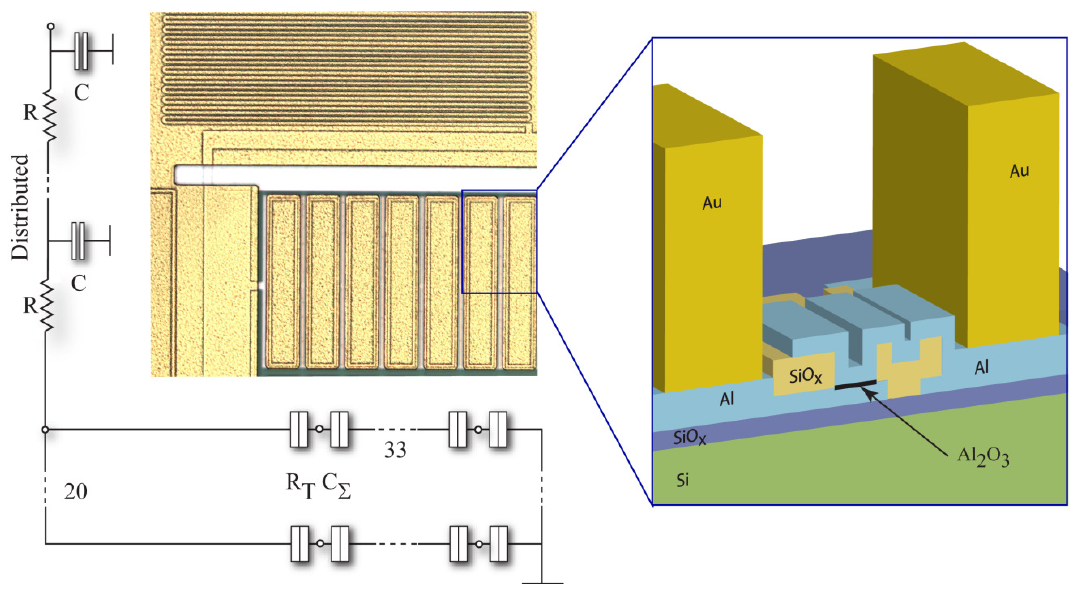}
\caption{
Optical micrograph of the CBT with equivalent circuit diagram (left) and schematic cross-section of the structure (right). The CBT is formed of $32\times20$ metallic islands of capacitance $C_\Sigma$ connected in an array by tunnel junctions of resistance $R_T$, as shown in the circuit diagram. Connection to the array is made via on-chip $RC$ filters comprising a meandering electrode sandwiched between large-area grounded metal films, separated by $250\,\mathrm{nm}$ $\mathrm{SiO_x}$. Each filter has a distributed resistance $R\approx 500\,\mathrm{\Omega}$ and capacitance $C\approx10\,\mathrm{pF}$. The schematic cross-section (right) shows one tunnel junction connecting two islands, with Au thermalisation blocks on top. \label{fig1}}
\end{figure}

The structure of the CBTs studied here is shown in Fig.~\ref{fig1}. Devices are fabricated using an ex situ tunnel junction process\cite{Prunnila2010} that provides excellent tunnel junction uniformity\cite{Hahtela2013}, and has also been used to fabricate superconducting qubits\cite{Gunnarsson2013}. In this work the CBTs are made from Al films with a thickness $250\,\mathrm{nm}$ and an isolating layer of $250\,\mathrm{nm}$ $\mathrm{SiO_x}$ deposited by PECVD. The tunnel junctions have a nominal diameter $0.8\,\mathrm{\mu m}$ and a resistivity $\sim 10\,\mathrm{k\Omega\,\mu m^2}$. The substrate is undoped Si with $300\,\mathrm{nm}$ thermal oxide on the surface.

Efficient thermalisation of electrons and phonons in the metallic islands of the CBT is critical for reaching low electron temperatures~\cite{Meschke2004,Feshchenko2013}. The electron-phonon heat flow $P_\mathrm{ep}$ is described by the material-dependent electron-phonon coupling constant $\Sigma$ and the volume of the metallic island $\Omega$,
\begin{equation}
P_\mathrm{ep} = \Sigma \Omega \left(\Te^5 - \Tp^5 \right)
\end{equation}
where $\Te$ is the electron temperature and $\Tp$ is the phonon temperature\cite{Wellstood1994}. To minimise $\Te$, the island volume should be large and the material chosen to maximise $\Sigma$. The island thickness achievable with the ex situ tunnel junction process, or other deposition techniques used for tunnel junction devices, is typically up to $1\,\mathrm{\mu m}$. Thicker films suffer from stress build-up, causing poor adhesion between the film and the substrate. This is a severe problem at mK temperatures where poor adhesion can lead to poor thermalisation and even mechanical failure due to thermal motion during cool-down. We avoid these problems by using a combination of the ex situ process followed by masked electroplating of Au on top of the CBT islands\cite{Xu2013}, which we refer to as thermalisation blocks. Electroplating can produce $\sim\,10\,\mathrm{\mu m}$ thick, low stress films, and here we choose a nominal thickness of $5\,\mathrm{\mu m}$ for the thermalisation blocks. The effective electron-phonon coupling in these islands, with nominal volume $5 \times 205 \times 38.5\,\mathrm{\mu m^3}$ and a high coupling constant\cite{Echternach1992} in Au $\Sigma = 2.4 \times 10^9\,\mathrm{W\,K^{-5}\,m^{-3}}$,  is estimated to be more than two orders of magnitude larger than in previous CBTs fabricated using the ex situ junction process\cite{Roschier2012}.

In addition to the thermalisation of the CBT itself it is important to thermalise the incoming leads through robust thermal anchoring and heavy electromagnetic filtering\cite{Bladh2003}. We improve the chain of thermalisation and filtering by including on-chip resistive meander structures in line with all electrical contacts. These form a distributed resistive-capacitive chain with a cut-off frequency of $\approx 40\,\mathrm{MHz}$. Similar filters based on a large area capacitor and tunnel junctions in series have previously been incorporated in a CBT \cite{Roschier2012}.

\begin{figure}
\includegraphics{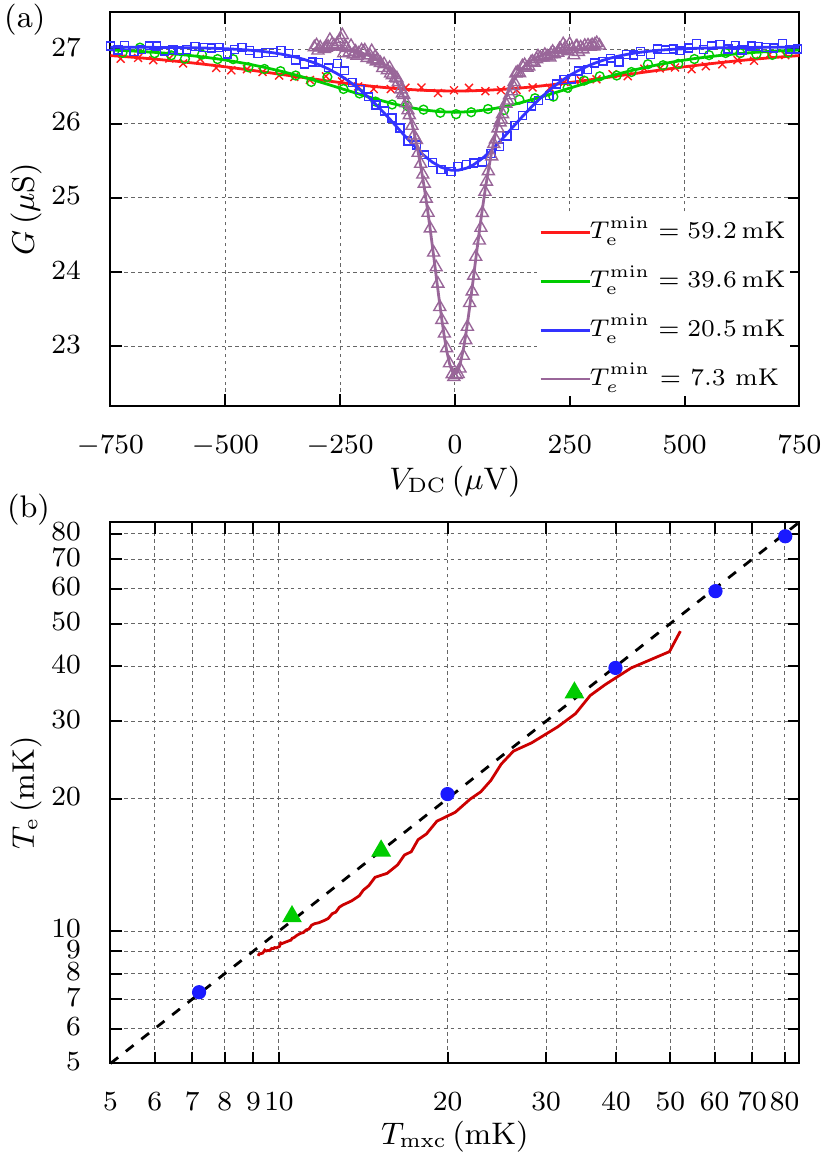}
\caption{CBT behaviour between $80\mK$ and $7\mK$ combining measurements made in two dilution refrigerators. (a) CBT conductance $G$ versus measured bias voltage $\Vdc$ at four temperatures. Symbols show measured values and lines show best fits to the calculated ideal conductance (see text for details of fitting). The warmest measurements (crosses, circles and squares) are fitted simultaneously to calibrate the sensor, giving $C_{\Sigma} = 236.6\,\mathrm{fF}$ and $R_T = 22.42\,\mathrm{k\Omega}$. The coldest measurement (triangles) was fitted using this calibration. The minimum electron temperatures $T_\mathrm{e}^\mathrm{min}$ are in close agreement with the refrigerator temperature measured by the $\mathrm{RuO_2}$ thermometer: $59.9\mK$, $40.1\mK$, $20.0\mK$, and $7.2\mK$ respectively. (b) CBT electron temperature $T_\mathrm{e}$ vs refrigerator temperature $T_\mathrm{mxc}$. Symbols show $T_\mathrm{e}^\mathrm{min}$ from fits to conductance dips measured in the cryogen-free refrigerator (circles) and the custom refrigerator (triangles). The solid line shows $T_\mathrm{e}$ determined by monitoring the conductance at $\Vdc=0$ as the cryo-free fridge cooled over $35\,\mathrm{min}$, showing that the CBT has a stronger thermal link to the refrigerator than the $\mathrm{RuO_2}$ thermometer, leading to the thermal lag ($\Tmxc \ge \Te$) during this time.
\label{fig2}}
\end{figure}

Figure~\ref{fig2} shows the behaviour of a CBT, fabricated using the process described above, focusing on temperatures between $7\mK$ and $80\mK$. The sensor was measured in a commercial cryogen-free dilution refrigerator\footnote{Bluefors Cryogenics LD250} with a base temperature $\approx 7\mK$ and later in a custom dilution refrigerator manufactured at Lancaster University\cite{Bradley1994} with a base temperature $\approx 2.5\mK$ (see below). The conductance of the CBT was measured in a current-driven four-wire configuration, with the drive current and voltage amplification provided by an Aivon PA10 amplifier. A small AC excitation (typically $5\,\mathrm{pA} \le \Iac \le 100\,\mathrm{pA}$) was added to the DC bias $\Idc$, allowing the differential conductance $G$ to be measured with a lock-in amplifier.

As shown in Fig.~\ref{fig2}(a), the CBT conductance dips around zero bias, and the dip becomes deeper and narrower at lower temperatures. Its full-width at half maximum is related to temperature by $V_{1/2} \approx 5.439 N k_B T/e$, where $N$ is the number of tunnel junctions in series~\cite{Pekola1994}. This does not account for self-heating in the sensor and so is only applicable when $T = \Te = \Tp$. A more practical parameter to determine temperature is the zero-bias conductance $G_0$, which has an approximate analytic relation to temperature~\cite{Farhangfar1997}
\begin{equation} \label{ggt_analytic}
G_0 \approx G_T\left( 1 - 1/6 u_N -1/60 u_N^2 + 1/630 u_N^3 \right)
\end{equation}
where $u_N \equiv E_C/k_B T$ is dimensionless inverse temperature, $E_C \equiv [(N-1)/N] e^2/C_\Sigma$ is the charging energy of the system, $C_\Sigma$ is the total capacitance of each island, and $G_T$ is the asymptotic conductance. When $u_n < 2.5$ the temperature measurement error is $<2.5$\%\cite{Feshchenko2013}. Thus, if $C_\Sigma$ and $G_T$ are known it is possible to determine $T$ by measuring only $G_0$. The most complete method to determine temperature is using a full tunnelling model to calculate $G(\Vdc)$ numerically~\cite{Pekola1994}. We use this last approach to find $C_\Sigma$ and $G_T$ for the device in order to then determine the temperature.

Numerical calculations of conductance are made using an algorithm derived from the free, open-source library pyCBT~\footnote{L. R. Roschier (Aivon Oy) et al. The pyCBT Library (2015). \href{https://github.com/AivonOy/pyCBT}{https://github.com/AivonOy/pyCBT}}. In addition, we account for overheating in the sensor by predicting the electron temperature $T_e$ in the islands using a model for the heat flow into each island,
\begin{equation}
P = \frac{\Vdc^2}{R_T} - \Sigma \Omega \left(\Te^5 - T_p^5\right) + P_0
\label{equ_therm}\end{equation}
where the first term is Joule heating at tunnel junctions of resistance $R_T$, the second term is heat flow to phonons, and $P_0$ accounts for parasitic heating. For a given phonon temperature, the minimum electron temperature $\Te^\mathrm{min}$ is found at $V_{DC}=0$. If $P_0$ is small, then $\Te^\mathrm{min} \approx T_p$.

Figure~\ref{fig2}(a) shows the result of fitting the calculated $G(\Vdc)$ simultaneously to three measurements made between $20\mK$ and $60\mK$. The fit parameters are $G_T$, $C_\Sigma$, and $T_p$ for each measurement. The fit was found to be insensitive to the value of $P_0$, and hence parasitic heating is assumed to be negligible at these temperatures. Having calibrated the CBT, the fitted $C_\Sigma$ was used to fit further measurements with $T_p$ and $G_T$ as the free parameters. An example is given in Fig.~\ref{fig2}(a), where the electron temperature is found to be $7.3\mK$ at a refrigerator temperature of $\Tmxc = 7.2\mK$, here measured by a calibrated $\mathrm{RuO_2}$ resistor\footnote{Sensor model RU-1000-BF0.007 supplied and calibrated by Bluefors Cryogenics.}.

Figure~\ref{fig2}(b) shows that the electron temperature measured by the CBT is in close agreement with the refrigerator temperature $\Tmxc$ in both refrigerators between $7\mK$ and $80\mK$. In the custom refrigerator, $\Tmxc$ is measured using a conventional Vibrating Wire Resonator viscometer (VWR) immersed in the saturated dilute phase of the $^3$He/$^4$He refrigerant in the mixing chamber\cite{Zeegers1991,Pentti2011}. For this set of measurements, in both refrigerators the CBT is in vacuum and housed in a gold-plated copper package (Aivon Oy SH-1) that is attached to the mixing chamber plate. The package includes RC filters with a cut-off frequency $\approx 300\,\mathrm{kHz}$. Electrical contacts to the CBT are thermalised in additional cold $RC$ filters potted in Eccosorb CR 124 (Aivon Oy `Therma'), which are also attached to the mixing chamber plate. The filters have a cut-off frequency $\approx 15\,\mathrm{kHz}$.

\begin{figure*}
\includegraphics{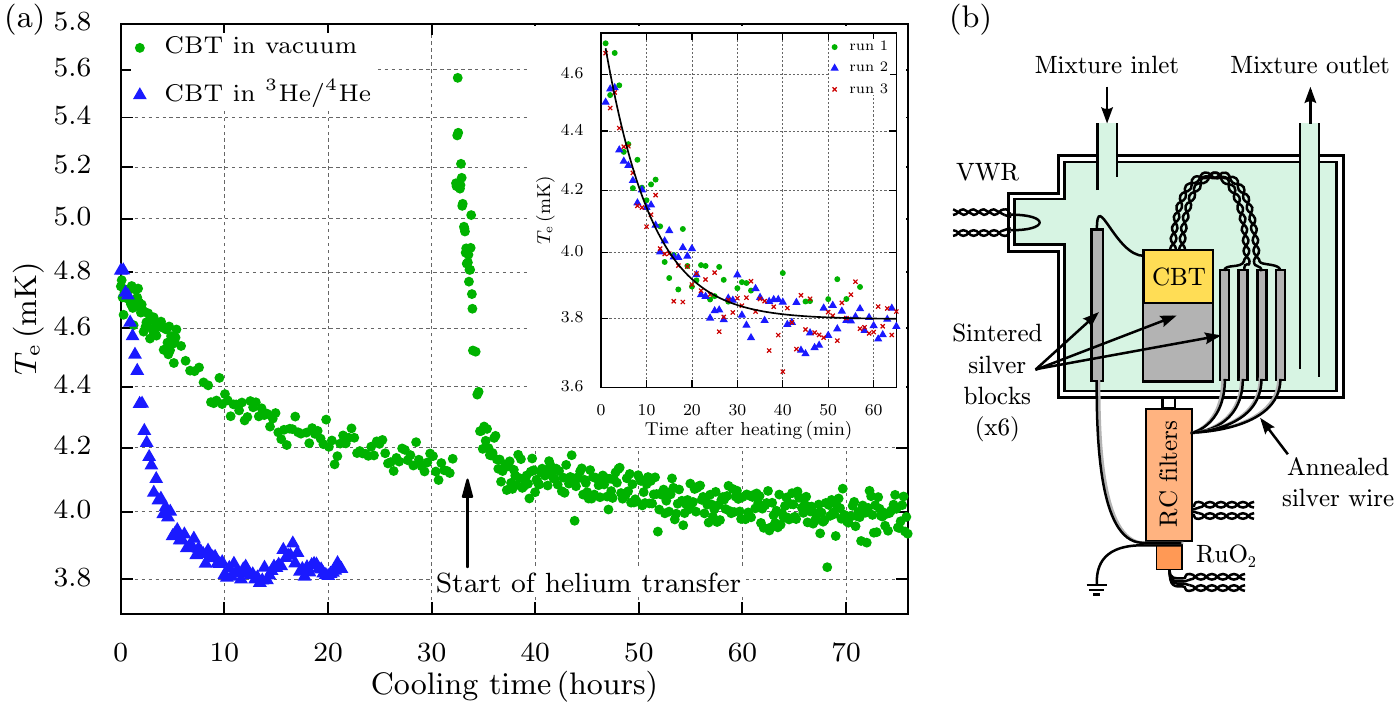}
\caption{Thermalisation of two CBTs as a function of time at a refrigerator temperature of $T_\mathrm{mxc} \le 2.8\mK$. (a) shows cooling of one CBT in vacuum (circles) and one immersed in the $^3$He/$^4$He refrigerant of the dilution refrigerator (triangles). In both cases, the CBTs are cooling after being warmed above $10\mK$ by temporarily increasing the refrigerant temperature. The CBT in vacuum is extremely slow to thermalise. By comparison, the CBT immersed in $^3$He/$^4$He thermalises significantly faster. The inset shows cooling of the immersed CBT after it has been heated by a large DC drive current ($50\,\mathrm{nA}$, $40\,\mathrm{nA}$ and $30\,\mathrm{nA}$ for run 1, 2 and 3 respectively). Fitting to an exponential decay (solid line) yields a time constant of $570\,\mathrm{s}$ and a saturation temperature of $3.8\mK$.(b) shows schematic of the immersion cell.\label{fig3}}
\end{figure*}

Subsequently we investigated the behaviour of the CBT below $7\mK$ in the custom dilution refrigerator. Figure~\ref{fig3}(a) shows the electron temperature of the sensor, when in vacuum, gradually cooling below $4\mK$ while the refrigerator was held at $\Tmxc \lesssim 2.8\mK$. Here $T_e$ was determined from the value of $G_0$, which was found by measuring $G$ over a small range of $\Vdc$ ($\approx 30\,\mathrm{\mu V}$) close to $\Vdc=0$. We observe an extremely long equilibration time (over $3\,\mathrm{days}$) but a rapid cooling of the CBT following a heating event (a refill of the liquid helium bath that briefly heated the CBT above $5.5\mK$ and the fridge above $3.5\mK$). This suggests that the thermal contact between the CBT and the refrigerator is relatively strong, and that its slow cooling is due to heat leak from an external warm object.

A second CBT was immersed in the $^3$He/$^4$He refrigerant of the custom dilution refrigerator to improve thermal coupling and better isolate external sources of heating. A schematic of the immersion cell is shown in the Fig.~\ref{fig3}(b). Sintered silver blocks increase the thermal contact between the refrigerant and the CBT and contact wires. Several sinters was attached to the sensor package, to each of the four measurement wires, and a grounding wire for the package and the $RC$ filters. The immersed CBT was found to equilibrate much faster, as shown in Fig.~\ref{fig3}(a), reaching $\Te = 3.8\mK$ at $\Tmxc = 2.7\mK$.

In order to study the CBT at $|\Vdc| \gg 0$, the time needed for the sensor to reach thermal equilibrium after a change of Joule heating needs to be known. The inset to Fig.~\ref{fig3}(a) shows the relaxation of $\Te$ after the CBT has been heated by a large drive current for long enough to reach thermal equilibrium ($>30\,\mathrm{min}$). The subsequent value of $\Te$ is measured by scanning close to $\Vdc=0$, where Joule heating should be negligible. The relaxation of $\Te$ is found to have a time constant of $570\,\mathrm{s}$.

\begin{figure}
\includegraphics{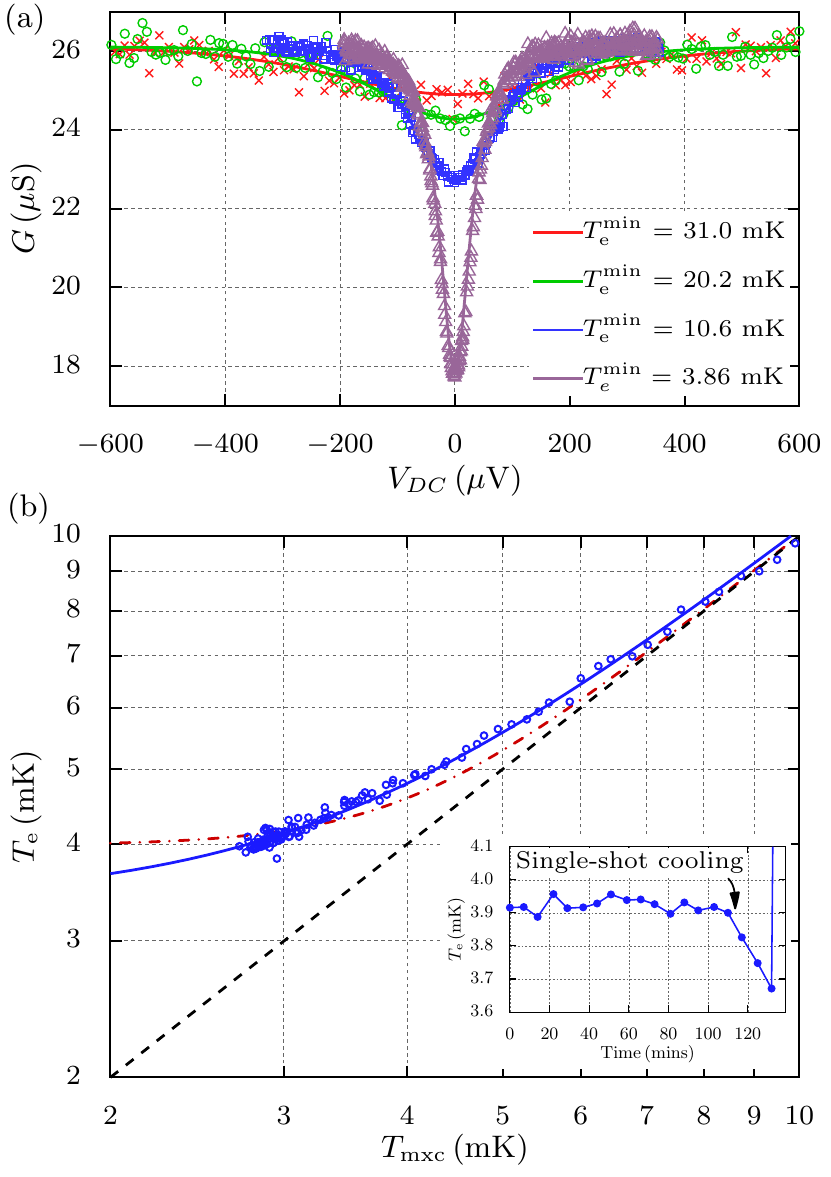}
\caption{Characteristics of a CBT immersed in $^3$He/$^4$He refrigerant. (a) Fitting to the warmest three measurements gives $C_{\Sigma} = 209.5\,\mathrm{fF}$ and $R_T = 23.21\,\mathrm{k\Omega}$. The fitted minimum electron temperatures $T_\mathrm{e}^\mathrm{min}$ for the warmest three curves are in reasonable agreement with the refrigerator temperature as measured by the VWR thermometer: $29.4\mK$, $19.0\mK$, $10.5\mK$ respectively. (b) shows measured electron temperature in the CBT as the refrigerator cooled steadily from $10\mK$ to $2.7\mK$ over a period of 12 hours. The solid line shows a fit of the form $T_\mathrm{e}^x = T_\mathrm{mxc}^x + c$, yielding an exponent $x = 2.7$. The dot-dashed line shows a best fit of $T_\mathrm{e}^5 = T_\mathrm{mxc}^5 + c$. The inset to (b) shows the measured $T_e$ as the refrigerator was temporarily cooled in single-shot mode to $2.2\mK$, reaching a lowest $\Te$ below $3.7\mK$.\label{fig4}}
\end{figure}

Figure~\ref{fig4}(a) shows the calibration of the immersed sensor. The three warmest measurements were fitted simultaneously to determine $C_\Sigma=209.5\,\mathrm{fF}$ and $R_T=23.21\,\mathrm{k\Omega}$. The fitted temperatures agree with $\Tmxc$  to within 6\%. Given the agreement between the fitted $\Te^\mathrm{min}$ and $\Tmxc$, we can assume that parasitic heating is still negligible down to $10\mK$.

The coldest measurement in Fig.~\ref{fig4}(a) was fitted using the above values, yielding a minimum electron temperature of $3.9\mK$. This measurement was made over a period of $7\,\mathrm{hours}$ to ensure that the CBT was in thermal equilibrium at each value of $\Vdc$. At these temperatures the parasitic heating of the CBT is now significant and $\Te^\mathrm{min}$ does not match the refrigerator temperature of $\Tmxc = 2.7\mK$. In order to fit this conductance dip the thermal model, Eq.~\ref{equ_therm}, was used with $\Tp = \Tmxc$, and with the parasitic heating $P_0$ and the electron-phonon coupling $\Sigma\Omega$ as free parameters. 

The overheating of the sensor at $\Vdc=0$ constrains the product $P_0\Sigma\Omega$; however, the parasitic heating is not large enough to reliably separate $P_0$ and $\Sigma\Omega$ in the fit. Qualitatively, the fits suggest that $P_0 \ge 300\,\mathrm{aW}$ per island and $\Sigma\Omega$ is at least four times larger than expected from the nominal size of the thermalisation blocks and the literature value of $\Sigma$ for Au\cite{Echternach1992}. It was not possible to determine an upper bound on $P_0$. It is worth noting that the power required to measure the CBT conductance ($< 1\,\mathrm{aW}$ per island due to Joule heating from $\Iac$) is much lower than our estimate of $P_0$. As such, we believe that CBTs of this type can be operated at still lower temperatures by reducing parasitic heating.

Figure~\ref{fig4}(b) shows how the CBT electron temperature diverges from the refrigerator temperature below $\approx 7\mK$. Here the value of $\Te$ is found by measuring $G_0$ close to $\Vdc=0$ and so Joule heating can be neglected. The functional form of $\Te$ vs $\Tmxc$ should have the same temperature dependence as the dominant thermalisation mechanism, i.e. $T^5$ for electron-phonon coupling. However, other power laws have been observed\cite{Casparis2012}. Here we find that the best fit of $\Te^x = \Tmxc^x + c$ gives $x = 2.7$ and a saturated $\Te$ of $c^{1/x} = 3.4\mK$. The fitted exponent $x$ cannot be confirmed by measurements of the conductance dip because the overheating is still relatively weak even at the lowest temperatures. We find that a thermal model with a $T^3$ thermalisation term fits the data equally well as a model using $T^5$. In order to understand the thermalisation mechanism in more detail, this sensor would need to be cooled closer to $1\,\mathrm{mK}$.

In conclusion, CBTs of the structure described here have been shown to operate as reliable primary thermometers of electron temperature down to $3.7\mK$. The large thermalisation blocks incorporated in the device, and a relatively low level of parasitic heating, ensure that the electron subsystem in the sensor is well coupled to the phonon subsystem down to $\approx 7\mK$. An immersion cell was shown to improve thermal coupling between a CBT and a dilution refrigerator. This allowed the onset of overheating to be observed below $7\mK$, and while the presence of overheating could be seen, the effect was sufficiently weak that the sensor will need to be cooled further to fully characterise the thermalisation mechanisms.

\section*{Acknowledgements}

We would like to thank Jukka Pekola and Matthias Meschke for useful discussions. This research is supported by the U.K. EPSRC (EP/K01675X/1 and EP/L000016/1), the European FP7 Programme MICROKELVIN (project No. 228464), Tekes project FinCryo (grant No. 220/31/2010) Academy of Finland (project No. 252598) and the Royal Society. JRP acknowledges support of the People Programme (Marie Curie Actions) of the European FP7 Programme under REA grant agreement 618450. YuAP acknowledges support by the Royal Society and the Wolfson Foundation.

\flushleft

\end{document}